\documentclass[cits]{PoS}
\usepackage{subfigure,graphicx}
\usepackage{amsmath}

\title{Velocity renormalization in graphene from lattice Monte Carlo}
\ShortTitle{Velocity renormalization in graphene from lattice Monte Carlo}

\author{
Joaqu\'in E. Drut$^a$ and \speaker{Timo A. L\"ahde}$^b$ \\ 
\llap{$^a$}Department of Physics and Astronomy, University of North Carolina, Chapel Hill,
North Carolina 27599-3255, USA \\
\llap{$^b$}Institute~for~Advanced~Simulation, Institut~f\"{u}r~Kernphysik, and 
J\"{u}lich~Center~for~Hadron~Physics,~Forschungszentrum~J\"{u}lich, D--52425~J\"{u}lich, Germany \\
E-mail: \email{drut@email.unc.edu}, \email{t.laehde@fz-juelich.de}
}

\abstract{
We compute the Fermi velocity of the Dirac quasiparticles in
clean graphene at the charge neutrality point for strong Coulomb coupling $\alpha_g^{}$. We perform a Lattice Monte Carlo
calculation within the low-energy Dirac theory, which includes an instantaneous, long-range Coulomb interaction.
We find a renormalized Fermi velocity $v_{FR}^{} > v_F^{}$, where $v_F^{} \simeq c/300$. Our results are consistent with a 
momentum-independent $v_{FR}^{}$ which increases approximately linearly with $\alpha_g^{}$, although
a logarithmic running with momentum cannot be excluded at present. At the predicted critical coupling 
$\alpha_{gc}^{}$ for the semimetal-insulator transition due to excitonic pair formation, we find $v_{FR}^{}/v_F^{} \simeq 3.3$, which we
discuss in light of experimental findings for $v_{FR}^{}/v_F^{}$ at the charge neutrality point in ultra-clean suspended graphene.
}


\FullConference{31st International Symposium on Lattice Field Theory LATTICE 2013 \\
		 July 29 -- August 3, 2013\\
		 Mainz, Germany}

\begin{document}

\newcommand{\be}{\begin{equation}}
\newcommand{\ee}{\end{equation}}
\newcommand{\ba}{\begin{eqnarray}}
\newcommand{\ea}{\end{eqnarray}}
\newcommand{\non}{\nonumber}
\newcommand{\sla}{\not\!}
\newcommand{\lang}{\left\langle}
\newcommand{\rang}{\right\rangle}
\newcommand{\al}{&\!\!\!}
\newcommand{\Lag}{\mathcal{L}}
\newcommand{\Amp}{\mathcal{A}}
\newcommand{\Tr}{\textrm{Tr}}
\newcommand{\re}{\textrm{Re}}
\newcommand{\order}[2][M_\pi]{\mathcal{O}(#1^{#2})}

\newcommand{\hcchicga}[1]{h_c'\to\chi_{c#1}\gamma}
\newcommand{\chichcga}[1]{\chi_{c#1}'\to h_c\gamma}
\newcommand{\Mo}[1]{ \overset{_\circ}{M}_{#1} }

%%%%%%%%%%%%%%%%%%%%%%%%%%%%%%%%%%%%%%%%%%%%%%%%%%%%%%%%%%%%%%%%%%%%%

\section{Introduction}

Graphene, a two-dimensional membrane of carbon atoms with unique electronic properties, 
is characterized by a low-energy spectrum of Dirac quasiparticles, with a Fermi velocity $v_F^{} \simeq 1/300$ of the speed
of light in vacuum~\cite{graphene,review_general}. As the strength of the Coulomb interaction between the quasiparticles is controlled by 
$\alpha_g^{} \equiv e^2/(4\pi\epsilon v_F^{}) \simeq 2.2/\epsilon$, the role of
interactions can be enhanced to the point that graphene may resemble Quantum Electrodynamics
(QED) in a strongly coupled regime~\cite{review_strong}. In particular, the unscreened, long-range Coulomb interaction in graphene leads to
non-trivial velocity renormalization effects. At weak coupling, a logarithmic running 
$v_{F}^{}(n)/v_F^{}(n_0^{}) = 1 + (\alpha_g^{}/4)\ln(n_0^{}/n)$ with carrier density $n$ is found~\cite{weak}, such that
$v_{FR}^{}$ is expected to become large in the vicinity of the Dirac point ($n = 0$).
On the experimental side, logarithmic velocity renormalization is most prominent in
ultra-clean suspended graphene~\cite{Elias_susp} and on boron nitride (BN) substrates~\cite{BoronNitride},
with $v_{FR}^{}/v_F^{} \simeq 2 - 3$ in the former case, where $\epsilon = 1$.

Electron-electron interactions may also trigger a semimetal-insulator transition due to
excitonic pairing of quasiparticles and holes at a critical coupling $\alpha_{gc}^{}$. 
For graphene, Lattice Monte Carlo (LMC) has been applied to the Dirac theory
using a contact Thirring interaction~\cite{Hands_graphene} and a long-range Coulomb 
interaction~\cite{DL_graphene, Bui1,Onogi}, and to the tight-binding Hamiltonian using interactions of the
Hubbard~\cite{Hubbard1,Hubbard2,Hubbard3} and Coulomb~\cite{Coulomb_hex,Bui2,Bui3} types.
For the Dirac theory, $\alpha_{gc}^{} \simeq 1.11 \pm 0.06$ was found~\cite{DL_graphene},
to be compared with $\alpha_{gc}^{} \simeq 0.9 \pm 0.2$ in the tight-binding + Coulomb approach~\cite{Bui2}.
While such a transition has not yet been observed,
the empirical $v_{FR}^{}/v_F^{} \simeq 2 - 3$ in suspended graphene is indicative of interaction-induced 
spectral changes~\cite{review_strong,Elias_susp}.

%%%%%%%%%%%%%%%%%%%%%%%%%%%%%%%%%%%%%%%%%%%%%%%%%%%%%%%%%%%%%%%%%%%%%

\section{Lattice Monte Carlo}

The LMC treatment
of graphene uses the linearized low-energy Hamiltonian~\cite{Semenoff, Carbotte} with an instantaneous 
Coulomb interaction, such that $A_\mu^{} \equiv (A_0^{},\vec 0)$.
This gives the Euclidean (continuum) action
\begin{equation}
S_E^{} \equiv \frac{1}{2g^2}\int d^3x \, dt \, (\partial_i^{} A_0^{})^2 + \sum_{a=1}^{N_f^{}} \int d^2x \, dt \,
\bar\psi_a^{} D[A_0^{}] \psi_a^{},
\label{Dirac}
\end{equation}
where $g^2 \equiv e^2/\epsilon$, with $\epsilon \equiv (1+\kappa)/2$ for a substrate with dielectric constant $\kappa$.
Here, $\psi_a^{}$ is a four-component Dirac field in $2+1$~dimensions with $\bar \psi \equiv \psi^\dagger \gamma_0^{}$, 
$A_0^{}$ is the gauge (photon) field in 
$3+1$~dimensions, and $N_f^{} = 2$ for a graphene monolayer. The Dirac operator is
\begin{equation}
D[A_0^{}] = \gamma_0^{} (\partial_0^{} + iA_0^{}) + v_F^{} \sum_{k=1}^2 \gamma_k^{}\partial_k^{} + m_0^{},
\end{equation}
where the $\gamma_\mu^{}$ satisfy the Euclidean Clifford algebra $\{\gamma_\mu^{},\gamma_\nu^{}\} = 2\delta_{\mu\nu}^{}$,
and the bare fermion mass $m_0^{}$ provides an infrared regulator for modes that would be massless when the
U($2N_f^{}$) chiral symmetry of Eq.~(\ref{Dirac}) is spontaneously broken.
The lattice version of Eq.~(\ref{Dirac}) is formulated in terms of ``staggered'' fermions,
{\it i.e.}~one-component Grassmann variables $\chi, \bar\chi$, which partially retain the U($2N_f^{}$) chiral symmetry of Eq.~(\ref{Dirac}) 
at finite lattice spacing (for staggered fermions in odd dimensions, see Ref.~\cite{staggered3}).
The fermionic part of Eq.~(\ref{Dirac}) is given for $N_f^{} = 2$ in terms of staggered fermions by
$\sum_{{\bf n},{\bf m}} {\bar\chi}_{\bf n}^{} \, K_{{\bf n},{\bf m}}^{}[\theta_0^{}] \,{\chi}_{\bf m}^{}$, 
where ${\bf n} \equiv (n_0^{},n_1^{},n_2^{}) = (t,x,y)$ and ${\bf m}$
denote lattice sites on a $2+1$ dimensional fermion ``brane''.
This square space-time lattice embedded in a cubic lattice should not be understood as a tight-binding theory.
The staggered Dirac operator is
\begin{align}
K_{{\bf n},{\bf m}}^{}[\theta_0^{}] = & \:
\frac{1}{2a}(\delta_{{\bf n}+{\bf e}_0^{},{\bf m}}^{} \, U_{0,\bf n}^{} - 
\delta_{{\bf n} - {\bf e}_0^{},{\bf m}}^{} \, U^{\dagger}_{0,\bf m})
+ \frac{\lambda}{2a} \sum_{i} 
\eta^i_{\bf n} (\delta_{{\bf n}+{\bf e}_i^{},{\bf m}}^{} - 
\delta_{{\bf n}-{\bf e}_i^{},{\bf m}}^{}) + m_0^{} \, \delta_{{\bf n},{\bf m}}^{},
\end{align}
where $\eta^1_{\bf n} = (-1)^{n_0^{}}$, $\eta^2_{\bf n} = (-1)^{n_0^{} + n_1^{}}$, with
${\bf e}_i^{}$ a unit vector in lattice direction $i$.
The invariance of Eq.~(\ref{Dirac}) under spatially uniform, time-dependent
gauge transformations is retained by the gauge links $U_0^{} \equiv \exp(i\theta_0^{})$.
We perform LMC calculations for $v_F^{} = 1$ (thus $g^2 \to g^2/v_F^{}$)
and $a/ a_x^{} = 1$, where $a \equiv a_t^{}$ is the temporal lattice spacing, and thus
$\lambda = 1$. At non-zero $\alpha_{g}^{}$, we have $\lambda_R^{} \equiv v_{FR}^{} \, (a/ a_x^{})_R^{}$ from which
$v_{FR}^{}/v_F^{}$ can be obtained, once the asymmetry $(a/ a_x^{})_R^{}$ is known.

We compute the renormalized $\lambda_R^{}$ and $m_R^{}$ from the staggered fermion propagator 
$C_f^{}(t,x,y) \equiv \langle \chi(t,x,y) \bar\chi(t_0^{},x_0^{},y_0^{}) \rangle
= \langle K_{{\bf n},{\bf n_0^{}}}^{-1} \rangle$ on an $N_x^2 \times N_t^{}$ space-time lattice with $N_{x,t}^{}/4$ integer. 
Here ${\bf n}_0^{}$ is an arbitrary point of reference, 
and the brackets denote an average over ensembles of gauge field configurations,
obtained as a function of $\beta \equiv v_F^{}/g^2 = 1/(4\pi\alpha_g^{})$ and $m_0^{}$,
with the Hybrid Monte Carlo algorithm.
From $C_f^{}(t,x,y)$, we form $C_{ft}^{}(t,p_1^{},p_2^{}) \equiv \sum_{x,y} \exp(-ip\cdot x) \, C_f^{}(t,x,y)$,
for momenta $p_i^{} = 2\pi n/N_x^{}$, with $n = -N_x^{}/4, \ldots, N_x^{}/4$,
$t = 0, \ldots, N_t^{}-1$, and summation over even lattice sites, defined by $(-1)^{t+x+y} = 1$.
The expression for the ``temporal correlator'' $C_{ft}^{R}$ with renormalized $m_R^{}$, $\lambda_R^{}$ and
wave function renormalization $Z_R^{}$ (for $a = 1$) is~\cite{Gockeler_long}
\begin{equation}
C_{ft}^{R}(t,p_1^{},p_2^{}) = Z_R^{} G_t^{}(p_1^{},p_2^{}),
\label{Cft_even}
\end{equation}
for $t = 0,2,\ldots,N_t^{}-2$, and
\begin{align}
C_{ft}^{R}(t,p_1^{},p_2^{}) = & -\frac{Z_R^{}}{2m_R^{}} \bigg[ \exp(iB_0^{}) \,G_{t+1}^{}(p_1^{},p_2^{})
- \exp(-iB_0^{}) \, G_{t-1}^{}(p_1^{},p_2^{}) \bigg], 
\label{Cft_odd}
\end{align}
for $t = 1,3,\ldots,N_t^{}-1$, with anti-periodic boundary conditions. The function $G_t^{}(p_1^{},p_2^{})$ is
\begin{align}
G_t^{}(p_1^{},p_2^{}) & \equiv \frac{N}{
%\cosh^2(\frac{\mu_t^{} N_t^{}}{2}) 
C^2(\mu_t^{}) - B^2(B_0^{})
%- \sin^2(\frac{B_0^{} N_t^{}}{2}) 
} 
%\left[ \cosh^2(\mu_t^{} N_t^{}/2) - \sin^2(B_0^{} N_t^{}/2) \right]^{-1}
\nonumber \\
& \times \bigg [ A(B_0^{}) C(\mu_t^{})
%\cos\left(\frac{B_0^{} N_t^{}}{2}\right) \cosh\left(\frac{\mu_t^{} N_t^{}}{2}\right)
\cos(B_0^{} t^\star_{}) \sinh(\mu_t^{} t^\star_{})
+ B(B_0^{}) D(\mu_t^{})
%\sin(B_0^{} N_t^{}/2) \sinh(\mu_t^{} N_t^{}/2)
\sin(B_0^{} t^\star_{}) \cosh(\mu_t^{} t^\star_{}) \nonumber \\
& \quad + i A(B_0^{}) C(\mu_t^{}) 
%\cos\left(\frac{B_0^{} N_t^{}}{2}\right) \cosh\left(\frac{\mu_t^{} N_t^{}}{2}\right)
\sin(B_0^{} t^\star_{}) \sinh(\mu_t^{} t^\star_{})
- i B(B_0^{}) D(\mu_t^{})
%\sin\left(\frac{B_0^{} N_t^{}}{2}\right) \sinh\left(\frac{\mu_t^{} N_t^{}}{2}\right)
\cos(B_0^{} t^\star_{}) \cosh(\mu_t^{} t^\star_{})
\bigg ],
\label{Gtform}
\end{align}
where $A(x) \equiv \cos(x N_t^{}/2)$, $B(x) \equiv \sin(x N_t^{}/2)$,
$C(x) \equiv \cosh(x N_t^{}/2)$, $D(x) \equiv \sinh(x N_t^{}/2)$,
$t^\star_{} \equiv N_t^{}/2 - t$, $N \equiv 2 m_R^{} / \sinh(2\mu_t^{})$, and
$\sinh^2(\mu_t^{}) \equiv m_R^2 + \lambda_R^2 \sin^2(p_1^{}) + \lambda_R^2 \sin^2(p_2^{})$.
%
%\begin{equation}
%%N \equiv 2 m_R^{} / \sinh(2\mu_t^{}) / 
%N \equiv \frac{2 m_R^{}}{\sinh(2\mu_t^{})} 
%\left[ \cosh^2(\mu_t^{} N_t^{}/2) - \sin^2(B_0^{} N_t^{}/2) \right]^{-1},
%%(f_3^2-f_2^2).
%\label{norm}
%\end{equation}
%%
%and the dispersion relation
%%
%\begin{align}
%\sinh^2(\mu_t^{}) & \equiv m_R^2 + \lambda_R^2 \sin^2(p_1^{}) + \lambda_R^2 \sin^2(p_2^{}).
%%\tilde m_t^2 & \equiv m_R^2 + \lambda_R^2 \sin^2(p_1^{}) + \lambda_R^2 \sin^2(p_2^{}),
%%\tilde m_t^2 & \equiv m_R^2 + \lambda_R^2 \sin^2(p_1^{}+B_1^{}) + \lambda_R^2 \sin^2(p_2^{}+B_2^{}),
%\label{disp_t}
%\end{align}
%
This expression for $C_{ft}^{R}$ includes a constant ``background field'' 
$B_0^{} \equiv \langle \theta_0^{} \rangle$, as $\langle \theta_0^{} \rangle \neq 0$ in a finite volume.
$B_0^{}$ is roughly bounded by $\pm \pi/N_t^{}$~\cite{Gockeler_long}, and
the imaginary part of $C_{ft}^{R}$ vanishes for $B_0^{} \to 0$.

\begin{figure}[t]
\centering
\subfigure[
$\lambda_R^{}$ as a function of $\beta \equiv 1/(4\pi\alpha_g^{})$ and $m_0^{}$, as obtained from $C_{ft}^{}$ and 
$C_{fx}^{}$ on $28^3 \times 8$ (upper panels), $32^3 \times 12$ and $32^4$ (lower panels) lattices. Note that the dependence
on $m_0^{}$ is negligible, and that the results obtained from $C_{ft}^{}$ and $C_{fx}^{}$ agree closely.]
{\raisebox{0.1cm}{\includegraphics[width = .50\columnwidth]{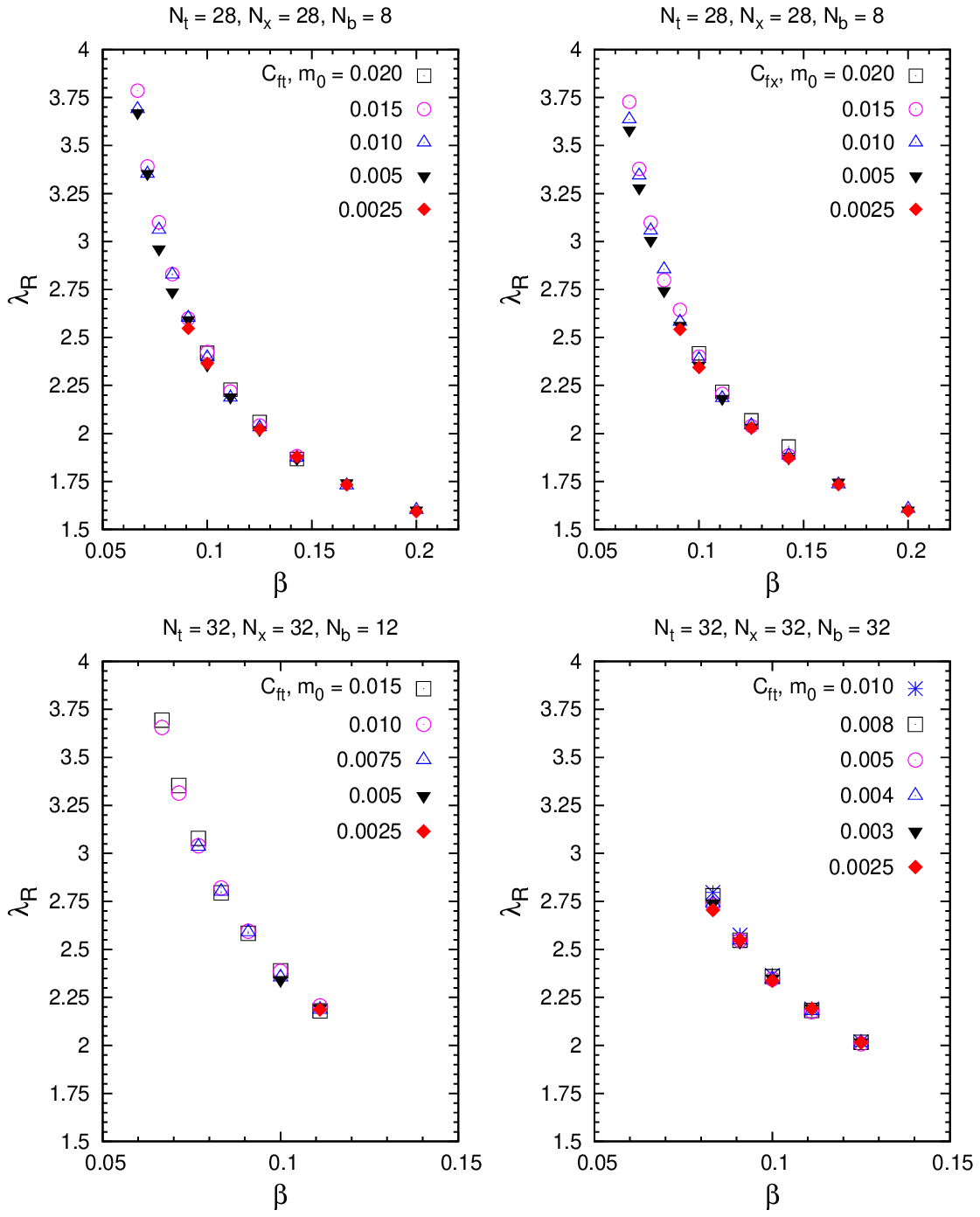}}\label{vf_lattice}}
\hspace{.02\columnwidth}
\subfigure[Left panel: $m_R^{}$ as a function of 
$\beta^{-1}_{} \equiv 4\pi\alpha_g^{}$ and $m_0^{}$,
obtained from $C_{ft}^{}$ (colored connected symbols) and $C_{fx}^{}$ (black unconnected symbols) 
on a $28^3 \times 8$ lattice. Right panel: $m_R^{}$ obtained from $C_{ft}^{}$ on $28^3 \times 8$
(colored connected symbols) and $32^3 \times 12$ lattices (black unconnected symbols).]
{\raisebox{-0.2cm}{\includegraphics[width = .46\columnwidth]{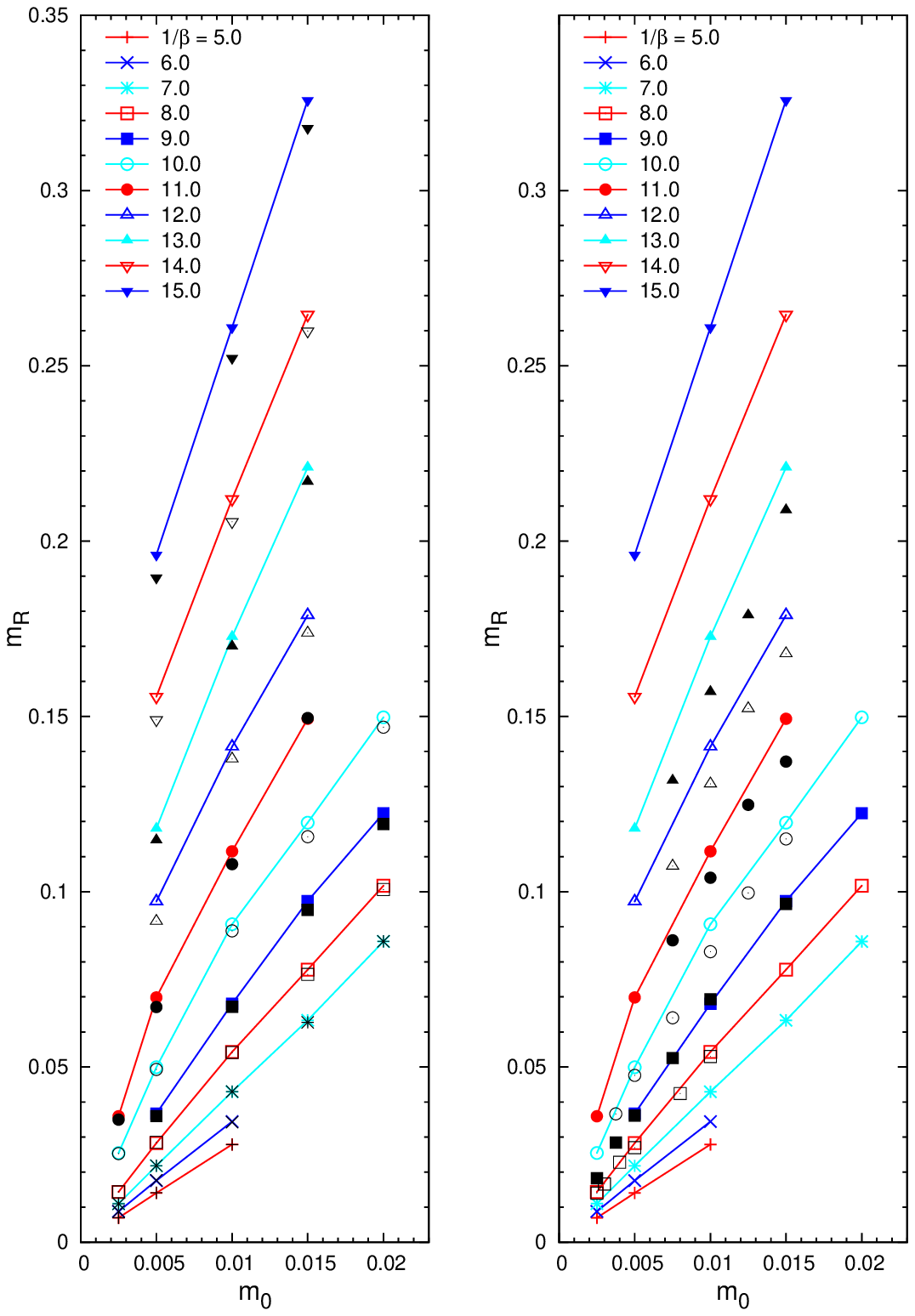}}\label{mr_lattice}}
\caption{LMC calculation of $\lambda_R^{}$ and $m_R^{}$ (for $a = 1$). 
All lines are intended as a guide to the eye, and statistical errors are comparable to the
size of the symbols.
}
\end{figure}

We also define the fermion ``spatial correlator''
$C_{fx}^{}(x,\omega,p_2^{}) \equiv \sum_{t,y} \exp(-ip\cdot x) \, C_f^{}(t,x,y)$,
for $\omega = 2\pi (n-1/2)/N_t^{}$ (due to the anti-periodic temporal boundary conditions), 
$n = -N_t^{}/4, \ldots, N_t^{}/4$,
spatial slices $x = 0,\ldots,N_x^{}-1$, and summation over even lattice sites.
The expression for $C_{fx}^{R}$ can be inferred from
$C_{ft}^{R}$. The function $G_x^{}(\omega,p_2^{})$ is obtained
from Eq.~(\ref{Gtform}) by first exchanging $\sin \leftrightarrow \cos$, followed by
$t \to x$, $N_t^{} \to N_x^{}$, $\mu_t^{} \to \mu_x^{}$ and $B_0^{} \to 0$. 
In addition, we have $m_R^{} \to m_R^{}/\lambda_R^{}$ in Eq.~(\ref{Cft_odd}),
$m_R^{} \to m_R^{}/\lambda_R^2$ in the expression for $N$, and
$\sinh^2(\mu_x^{}) \equiv m_R^2/\lambda_R^2 + \sin^2(\omega+B_0^{})/\lambda_R^2 + \sin^2(p_2^{})$.
Unlike Eqs.~(\ref{Cft_even}) and~(\ref{Cft_odd}), $C_{fx}^{R}$ is real-valued, with periodic boundary conditions.

%%%%%%%%%%%%%%%%%%%%%%%%%%%%%%%%%%%%%%%%%%%%%%%%%%%%%%%%%%%%%%%%%%%%%

\section{Results}

In Fig.~\ref{vf_lattice}, we show $\lambda_R^{}$ as obtained from a chi-square fit of Eqs.~(\ref{Cft_even}) and~(\ref{Cft_odd}) to LMC data. 
While Eq.~(\ref{Dirac}) is gauge invariant, $C_{ft}^{}$ and $C_{fx}^{}$ are not, 
and thus a gauge fixing condition is imposed on each configuration in order to obtain stable results.
For $C_{ft}^{}$, a correlated fit is performed for all $(t, p_1^{},p_2^{})$, and in the case of $C_{fx}^{}$ for all $(x, \omega,p_2^{})$.
The fitted parameters are $m_R^{}$, $\lambda_R^{}$, $B_0^{}$ and $Z_R^{}$. Our lattices
have $N_t^{} = N_x^{}$, and the length of the ``bulk'' dimension (where only the photons propagate) is denoted $N_b^{}$. We use the notation
$N_x^3 \times N_b^{}$, and simply $N_x^4$ when $N_b^{} = N_t^{} = N_x^{}$.
On the $28^3 \times 8$ lattice, LMC data is available for (inverse) lattice couplings 
$5.0 \leq \beta^{-1} \leq 15.0$, and for bare quasiparticle masses $0.0025 \leq m_0^{} \leq 0.020$, with slightly more
restrictive data sets on the $32^3 \times 12$ and $32^4$ lattices. We find that $\lambda_R^{}$ increases monotonically as a function of
$\alpha_g^{}$ from the non-interacting value of unity, with no appreciable differences between $\lambda_R^{}$ as obtained
from $C_{ft}^{}$ and $C_{fx}^{}$. We find the dependence on $m_0^{}$ to be almost negligible.
Finite size effects for $\lambda_R^{}$ are small, and the fitted values of $B_0^{}$ agree closely with $\langle \theta_0^{} \rangle$. 

In Fig.~\ref{mr_lattice}, we show the physical quasiparticle mass $m_R^{}$ as a function of $\beta^{-1}$ and $m_0^{}$, with emphasis on asymmetries
between the temporal and spatial correlations, and on finite size effects. As Eq.~(\ref{Dirac}) treats space and time asymmetrically, the spatial and 
temporal correlation lengths $\xi$ may exhibit unequal critical scaling, such that $\xi_s^{} \propto |\beta - \beta_c^{}|^{-\nu_s^{}}$ and 
$\xi_t^{} \propto |\beta - \beta_c^{}|^{-\nu_t^{}}$. The dynamical critical exponent $z \equiv \nu_t^{}/\nu_s^{}$ is an important characteristic of a 
quantum critical point (QCP), and implies that the dispersion relation is modified to $E \propto p^z_{}$.
At large $N_f^{}$, Ref.~\cite{Son} predicted $z \simeq 0.8$ for graphene in the strong-coupling limit.
However, arguments have also been put forward which indicate $z = 1$ for a QCP with $d < 4$ in theories with a long-range Coulomb
interaction~\cite{Herbut_z}. From Fig.~\ref{mr_lattice}, we find that the values of $m_R^{}$ obtained from $C_{ft}^{}$ and $C_{fx}^{}$
agree very closely for $\beta^{-1} \leq 11.0$, which is consistent with $z \simeq 1$. We also find no sign of non-linear dispersion.
A more accurate analysis is possible 
following Ref.~\cite{Hands_graphene}, in terms of the equation of state (EOS)
\begin{equation}
m_0^{} = A (\beta - \beta_c^{}) \, m_R^{(\delta \cdot \beta_m^{} - 1) / \nu_t^{}} + B \, m_R^{\delta \cdot \beta_m^{} / \nu_t^{}},
\label{EOS}
\end{equation}
for $m_R^{}$ computed from $C_{ft}^{}$, where $\delta$ and $\beta_m^{}$ 
are critical exponents characterizing the QCP at $\beta = \beta_c^{}$.
An analogous EOS with $\nu_t^{} \to \nu_s^{}$ can be given
for $m_R^{}$ as obtained from $C_{fx}^{}$. We also find from Fig.~\ref{mr_lattice} that finite size effects
are under control for $\beta^{-1} \leq 9.0$, but not for smaller $\beta$ (stronger coupling), especially in the region of the phase diagram where
a dynamically generated gap exists, and $\lim_{m_0^{} \to 0} m_R^{} \neq 0$. In principle, Eq.~(\ref{EOS}) can be used to compute
$z$ and $m_R^{}$ in the limit $m_0^{} \to 0$, {\it i.e.}~the gap in the quasiparticle spectrum. 
The lattice spacing asymmetry $(a/a_x^{})_R^{}$ could then be inferred by measuring the gap in terms of temporal and
spatial correlations. For reliable conclusions, such an EOS analysis should be combined with an extrapolation to 
infinite volume, similar to that of Ref.~\cite{Gockeler_EOS} for QED$_4^{}$. For the present analysis, 
we note from Fig.~\ref{mr_lattice} that the values of $m_R^{}$ computed from $C_{ft}^{}$ and $C_{fx}^{}$ differ by 
no more than $\simeq 10$\% at the smallest values of $\beta$. Assuming this trend persists in the limits of
infinite volume and vanishing $m_0^{}$, we find $1.0 \leq (a/a_x^{})_R^{} \leq 1.1$ over the range of $\alpha_g^{}$ studied.
In the absence of substantial indications for $(a/a_x^{})_R^{} \neq 1$, we take $\lambda_R^{}$ as a measure of $v_{FR}^{} / v_F^{}$.

%%%%%%%%%%%%%%%%%%%%%%%%%%%%%%%%%%%%%%%%%%%%%%%%%%%%%%%%%%%%%%%%%%%%%

\section{Conclusions}

\begin{figure}[t]
\centering
\includegraphics[width = .45\columnwidth]{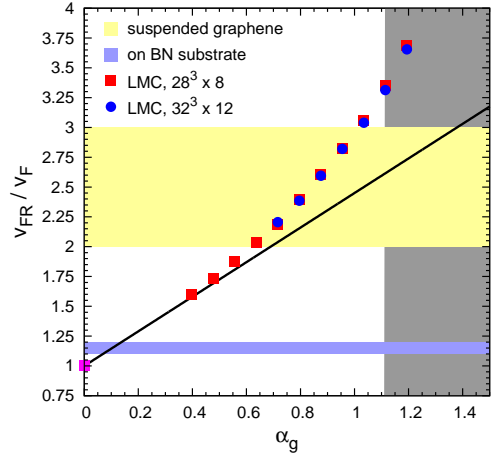}
\caption{Plot of $v_{FR}^{}(\alpha_g^{}) / v_F^{}$ for asymmetry $(a/a_x^{})_R^{} \simeq 1$. 
The black line is $v_{FR}^{} / v_F^{} = 1 + C\alpha_g^{}$,
and the predicted insulating phase occurs for $\alpha_g^{} > \alpha_{gc}^{}$ (gray shaded area). 
We find a linear dependence of $v_{FR}^{} / v_F^{}$ on $\alpha_g^{}$ up to $\alpha_g^{} \simeq 0.5$ (solid black line). 
Note that $v_{FR}^{}(\alpha_g^{} = 0) / v_F^{} \equiv 1$.
Horizontal bands indicate empirical data on $v_{FR}^{} (p = 0) / v_F^{}$ from
Ref.~\cite{Elias_susp} (suspended graphene) and Ref.~\cite{BoronNitride} (on BN substrate). 
Note that the expected logarithmic momentum-dependence of $v_{FR}^{} / v_F^{}$ cannot be resolved on present lattices.
\label{vf_comp}}
\end{figure}

%%%%%%%%%%%%%%%%%%%%%%%%%%%%%%%%%%%%%%%%%%

In Fig.~\ref{vf_comp}, we summarize our LMC results for $v_{FR}^{} / v_F^{}$ as a function of $\alpha_g^{} \equiv 1/(4\pi\beta)$, and compare with
available experimental data. Throughout our analysis, we have assumed that $v_{FR}^{}$ is constant, while in reality we should expect it
to run logarithmically with the momentum $p$ and diverge at the Dirac point, according to $v_{F}^{}(p)/v_F^{}(p_0^{}) = 1 + (\alpha_g^{}/4)\ln(p_0^{}/p)$, 
where $p_0^{}$ is the momentum scale at which $v_{FR}^{} = v_F^{}$. At present, we
cannot distinguish between this and the simpler expression $v_{FR}^{} / v_F^{} = 1 + C\alpha_g^{}$, and much
larger lattices appear to be required to detect a logarithmic running of the Fermi velocity with $p$. 
On the other hand, we find a pronounced dependence of
$v_{FR}^{} / v_F^{}$ on $\alpha_g^{}$. We find that $v_{FR}^{}$ increases linearly with $\alpha_g^{}$
from the non-interacting value $v_F^{}$ up to $\alpha_g^{} \simeq 0.5$,
above which the increase becomes more rapid. At the predicted critical coupling
$\alpha_{gc}^{} \simeq 1.1$, we find $v_{FR}^{}(\alpha_{gc}^{}) / v_F^{} \simeq 3.3$ within our present linearized treatment of the velocity
renormalization. Since all of the empirical $v_{FR}^{} (p = 0) / v_F^{}$ fall short of this, we find a plausible explanation for the
non-observation of excitonic insulating phases in graphene monolayers. We note that the result of Ref.~\cite{Elias_susp} is tantalizingly close to
$\simeq 3.3$, which suggests that further refinements in the quality of suspended graphene may suffice to trigger the excitonic instability.
It would be of interest to study the logarithmic running of $v_{FR}^{}$ with momentum on larger lattices.

\acknowledgments

We thank Lars Fritz, Simon Hands and Gerrit Schierholz for 
instructive discussions, and Jan Bsaisou, Dean Lee and Ulf-G.~Mei{\ss}ner
for a careful reading of the manuscript. T.~L. acknowledges 
financial support from the Helmholtz Association (contract VH-VI-417). This work was supported in part by an allocation of
computing time from the Ohio Supercomputer Center (OSC).

%%%%%%%%%%%%%%%%%%%%%%%

\end{document}